\documentclass[aps,prb,10pt,twocolumn]{revtex4-1}
\usepackage{bm}
\usepackage{amsmath}
\usepackage{amsfonts}
\usepackage{amssymb}
\usepackage{float}
\usepackage[final]{graphicx}
\usepackage[caption = false]{subfig}
\usepackage{dsfont}
\usepackage[colorlinks=true,linktocpage=true,breaklinks=true]{hyperref}

\begin{document}
\title{Elusive Dzyaloshinskii-Moriya interaction in Fe$_3$GeTe$_2$ monolayer}
\author{Slimane Laref$^1$}
\email{slimane.laref@kaust.edu.sa}

\author{Kyoung-Whan Kim$^2$}
\author{Aur\'elien Manchon$^{1,3}$}
\email{aurelien.manchon@kaust.edu.sa}
\affiliation{$^1$Physical Science and Engineering Division (PSE), King Abdullah University of Science and Technology (KAUST), Thuwal 23955-6900, Kingdom of Saudi Arabia\\
$^2$ Center for Spintronics, Korea Institute of Science and Technology, Seoul 02792, Korea\\
$^3$Aix Marseille Univ, CNRS, CINAM, Marseille, France.}

\begin{abstract}
Using symmetry analysis and density functional theory calculations, we uncover the nature of Dzyaloshinskii-Moriya interaction in Fe$_3$GeTe$_2$ monolayer. We show that while such an interaction might result in small distortion of the magnetic texture on the short range, on the longrange Dzyaloshinskii-Moriya interaction favors in-plane N\'eel spin-spirals along equivalent directions of the crystal structure. Whereas our results show that the observed N\'eel skyrmions cannot be explained by the Dzyaloshinskii-Moriya interaction at the monolayer level, they suggest that canted magnetic texture shall arise at the boundary of Fe$_3$GeTe$_2$ nanoflakes or nanoribbons and, most interestingly, that homochiral planar magnetic textures could be stabilized.
\end{abstract}
\maketitle

\paragraph*{Introduction - }
Magnetism in low dimensions has received renewed interest in the past few years with the experimental observation of remnant magnetization in two-dimensional van der Waals materials such as CrI$_3$ \cite{Huang2017b}, VTe$_2$ \cite{Li2018f}, CrTe$_2$ \cite{Sun2019}, and Fe$_3$GeTe$_2$ \cite{Deng2018}. The emergence of robust magnetic order at room temperature is appealing for spintronics applications, and among the ever-increasing family of candidate materials Fe$_3$GeTe$_2$ stands out as a solid paradigm \cite{Deng2018,Zhang2020a,Park2020}. As a matter of fact, this material hosts interesting promises: spin-orbit torque \cite{Alghamdi2019,Wang2019c} and anomalous Nernst effect \cite{Xu2019,Fang2019} have been observed in bilayer heterostructures, and magnetoresistance has been reported in spin-valves \cite{Wang2018c,Albarakati2019}. \par

Besides these experimental achievements, the recent reports of magnetic skyrmions and other chiral textures in thick Fe$_3$GeTe$_2$ layers \cite{Wang2019d,Ding2019,Park2019} are intriguing. As a matter of fact, stable and metastable chiral magnetic textures require the existence of an antisymmetric exchange interaction, called Dzyaloshinskii-Moriya interaction\cite{Dzyaloshinskii1957,Moriya1960}. This interaction only exists in materials lacking inversion symmetry and the specific structure of this interaction determines the nature of the chiral magnetic structures it can stabilize \cite{Nagaosa2013}. For instance, in magnetic multilayers the interfacial symmetry breaking promotes the onset of an interfacial Dzyaloshinskii-Moriya interaction of the form $E_{\rm DM}=D{\bf m}\cdot[({\bf z}\times{\bm\nabla})\times{\bf m}]$ that favors N\'eel skyrmions (e.g., see Ref. \onlinecite{Chen2015b}). Therefore, the observation of N\'eel-type skyrmions in thick Fe$_3$GeTe$_2$ layer\cite{Wang2019d,Ding2019,Park2019} is unexpected as the point group of Fe$_3$GeTe$_2$ monolayer prevents the onset of "interfacial" Dzyaloshinskii-Moriya interaction. However, the point group of Fe$_3$GeTe$_2$ monolayer does not entirely forbid the emergence of chiral effects. As pointed recently by Johansen et al.\cite{Johansen2019}, point group symmetry analysis shows that Fe$_3$GeTe$_2$ monolayer exhibits a damping-like spin-orbit torque, while the field-like torque is zero. Since Dzyaloshinskii-Moriya interaction and damping-like torque are related to each other \cite{Freimuth2014}, one can expect a non-vanishing Dzyaloshinskii-Moriya interaction but of completely different nature compared to the interfacial one.\par
\begin{figure}
\begin{center}
        \includegraphics[width=7cm]{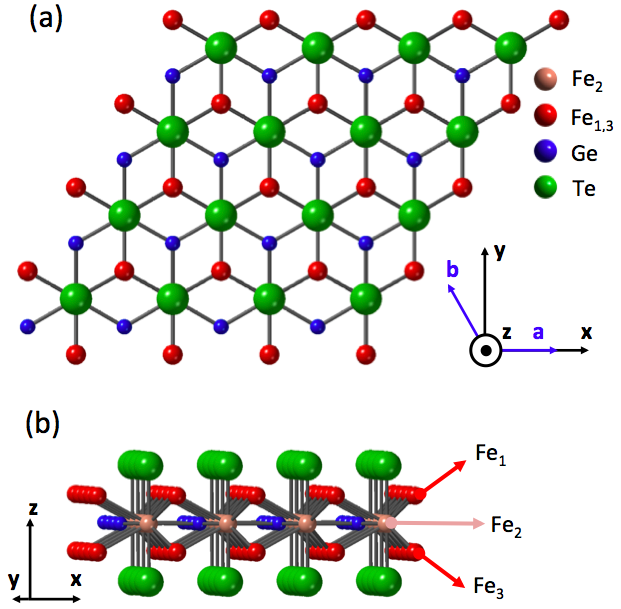}
      \caption{(Color online) (a) Top and (b) side view of Fe$_3$GeTe$_2$ monolayer. ($x$, $y$, $z$) are the cartesian coordinates and (${\bf a},{\bf b}$) are the equivalent crystallographic directions.\label{Fig0}}
\end{center}
\end{figure}
In this work, using symmetry analysis and density functional theory (DFT) calculations, we investigate the nature of Dzyaloshinskii-Moriya interaction in Fe$_3$GeTe$_2$ monolayer. We show that while such an interaction might result in small distortion of the magnetic texture on the short range, on the long-wavelength limit Dzyaloshinskii-Moriya interaction favors in-plane N\'eel spin-spirals along equivalent directions of the crystal structure. Whereas these results show that the observed N\'eel skyrmions cannot be explained by the Dzyaloshinskii-Moriya interaction at the monolayer level, they suggest that canted magnetic texture shall arise at the boundary of Fe$_3$GeTe$_2$ nanoflakes or nanoribbons and that homochiral planar magnetic textures can be stabilized.\par

\paragraph*{Long-wavelength behavior - }
Let us first consider the crystal structure of Fe$_3$GeTe$_2$ monolayer, depicted on Fig. \ref{Fig0}. The crystal adopts the point group $\bar{6}m1$ and can be seen as a stack of three Fe hexagonal lattices in A-B-A configuration. In the following the central Fe element is denoted Fe$_2$ and the Fe elements on the top and bottom planes are referred to as Fe$_{1,3}$, respectively. One can see by inspection on Fig. \ref{Fig0} that all the three inequivalent Fe elements are located in a chemical environment that lacks inversion symmetry. Therefore, one can expect each magnetic element to experience chiral effects such as Dzyaloshinskii-Moriya interaction and spin-orbit torques in the presence of spin-orbit coupling. However, Fe$_1$ and Fe$_3$ are mirror partners, i.e., related by mirror symmetry. Therefore, any chiral physical quantity on one element is opposite on the other element. In contrast, Fe$_2$ is located in the mirror plane of the crystal and therefore should experience such chiral effects.\par

As a matter of fact, a first indication of the existence of spin-orbitronics effects was provided by analyzing the point group of Fe$_3$GeTe$_2$ monolayer. For instance, applying these symmetries [improper six-fold rotation about (001), mirror symmetry normal to (110)] to the current-driven field response tensor \cite{Zelezny2017}, one obtains a vanishing field-like torque but an unusual non-zero damping-like torque (see also Ref. \onlinecite{Johansen2019})
\begin{eqnarray}
{\bf T}_{\rm DL}=\eta{\bf m}\times[(m_yE_x+m_xE_y){\bf x}+(m_xE_x-m_yE_y){\bf y}].
\end{eqnarray}
In this expression, $\eta$ is the torque response coefficient and ${\bf E}$ is the applied electric field. This torque is particularly interesting as it behaves like a non-equilibrium anisotropy energy term \cite{Johansen2019}. Of major interest to the present work, one can show that {\em in the limit of small spatial gradients}, i.e., in the long-wavelength limit, the Dzyaloshinskii-Moriya tensor has the same symmetry as the damping-like torque response \cite{Freimuth2014}. In fact, defining the torque response tensor $\hat{\chi}_{\rm DL}$ as ${\bf T}_{\rm DL}=\hat{\chi}_{\rm DL}\cdot {\bf E}$ and the Dzyaloshinskii-Moriya tensor $\hat{D}$ as $E_{\rm DM}=\sum_{ij}D_{ij}{\bf e}_i\cdot({\bf m}\times\partial_j{\bf m})$, the linear response theory yields $\hat{\chi}_{\rm DL}\propto\hat{D}$. In other word, the torque tensor reads
\begin{equation}
\hat{\chi}_{\bar{6}m1}=\eta\left(\begin{matrix}
-m_zm_x&m_zm_y &0 \\
m_zm_y&m_zm_x&0\\
m_x^2-m_y^2 &-2m_ym_x&0\\
\end{matrix}\right)\propto\hat{D}_{\bar{6}m1}
\end{equation}
One can deduce the Dzyaloshinskii-Moriya energy,
\begin{eqnarray}\label{eq:dmilongrange}
E_{\rm DM}&=&D[-\partial_x(m_ym_x)+\frac{1}{2}\partial_y(m_x^2-m_y^2)].
\end{eqnarray}
Hence, since the Dzyaloshinskii-Moriya interaction is a total derivative, it does not stabilize chiral textures in the long-wavelength limit. However, one could wonder whether this interaction can stabilize magnetic twists at the edges of the magnetic layer, as discussed recently \cite{Raeliarijaona2018,Hals2017,Rohart2013}. To investigate this possibility, let us a magnetic ribbon with easy-plane anisotropy and embedded between two boundaries normal to the direction ${\bf n}$. The system is translationally invariant along ${\bf z}\times{\bf n}$ and therefore, spatial gradients are only allowed along ${\bf n}$, ${\bm\nabla}=\partial_{x_n}{\bf n}$, where $x_n$ is the coordinate along {\bf n}. In the bulk of the nanoribbon, the magnetization ${\bf m}$ minimizes the energy functional $W=A(\partial_{x_n}{\bf m})^2+K({\bf m}\cdot{\bf z})^2$, where $A$ is the exchange and $K$ the easy plane anisotropy. The general solution is ${\bf m}=\cos(ax_n+\phi){\bf n}+\sin(ax_n+\phi){\bf z}\times{\bf n}$, where ${\bf m}$ is expressed in the frame (${\bf n}$, ${\bf z}\times{\bf n}$, ${\bf z}$). The boundary condition reads \cite{Hals2017},
\begin{eqnarray}\label{eq:bc1}
2A({\bf n}\cdot{\bm\nabla}){\bf m}+{\bf m}\times({\bm\Gamma}_{\rm D}\times{\bf m})=0
\end{eqnarray}
where ${\bm\Gamma}_{\rm D}$ is the boundary-induced Dzyaloshinskii-Moriya field, defined ${\bm\Gamma}_{\rm D}=m_in_jD_{ijk}$, with $E_{\rm DM}=\sum_{ijk}D_{ijk}m_i\partial_jm_k$ being the Dzyaloshinskii-Moriya energy. Solving Eq. \eqref{eq:bc1} at the positions $x_n=x_1$ and $x_n=x_2$, we obtain two coupled equations
\begin{eqnarray}\label{eq:bc2}
\cos2(ax_{1,2}+\phi)=(2A/D)a,\\
\cos2(ax_2+\phi)=(2A/D)a,
\end{eqnarray}
yielding the solution, $\sin(a(x_1+x_2)+2\phi)\sin(a(x_2-x_1))=0\Rightarrow a=n\pi/(x_2-x_1)$. However, the total energy of this spin spiral is $W=An^2\pi^2/L^2$, which is minimized for $n=0$. Therefore, the Dzyaloshinskii-Moriya interaction given by Eq. \eqref{eq:dmilongrange} does not favor chiral magnetic textures, even in the case of a planar ferromagnet. We emphasize that the present discussion only concerns the interaction derived in the limit of small spatial gradients. It does not address the possible existence of short-wavelength magnetic textures.

\begin{figure}
\begin{center}
        \includegraphics[width=8cm]{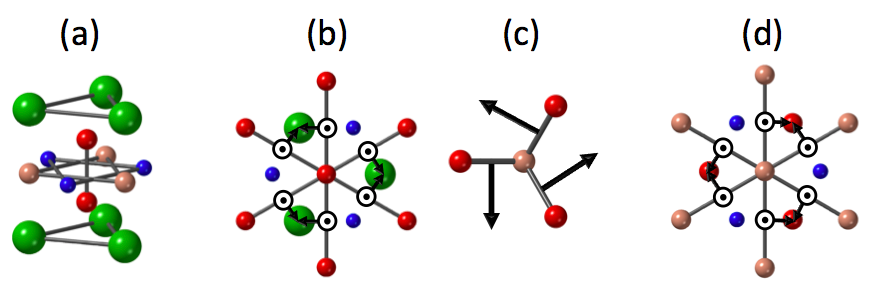}
      \caption{(Color online) Dzyaloshinskii-Moriya vector (black arrows) for various nearest neighbor interactions: (a) Fe$_1$-Fe$_3$, (b) Fe$_1$-Fe$_2$, (c) Fe$_1$-Fe$_1$ and (d) Fe$_2$-Fe$_2$. The chemical elements are designated by the same color code as in Fig. \ref{Fig0}. In these figures, we only represented the atoms that contribute to defining the local symmetry of the environment and removed the other elements for better clarity. \label{Fig1}}
\end{center}
\end{figure}

\paragraph*{Structural analysis - } Let us now take a different perspective and consider the atomistic Dzyaloshinskii-Moriya interaction between neighboring magnetic moments. The relevant pairs of neighboring moments are displayed on Fig. \ref{Fig1} together with the Dzyaloshinskii-Moriya vector ${\bf D}$, defined in the atomistic spin limit $E_{\rm DM}={\bf D}\cdot({\bf S}_1\times{\bf S}_2)$. The Dzyaloshinskii-Moriya vector is determined by Moriya's rules \cite{Moriya1960}. \par

We first consider the Fe$_1$-Fe$_3$ pair, located on each side of the (001) mirror plane [Fig. \ref{Fig1}(a)]. Since the axis on which this pair lies has a three-fold rotational symmetry, the Dzyaloshinskii-Moriya vector is necessarily along the axis (5th Moriya rule). But since this axis also possesses a mirror symmetry perpendicular to it, the Dzyaloshinskii-Moriya vector must be also perpendicular to the axis (1st Moriya rule). As a result, there is no Dzyaloshinskii-Moriya interaction between Fe$_1$ and Fe$_3$. Let us now consider the interaction between two Fe$_1$ (or, equivalently, two Fe$_3$) belonging to the same layer [Fig. \ref{Fig1}(b)]. Since a mirror plane passes perpendicularly through the center of the Fe$_1$-Fe$_1$ axis, the Dzyaloshinskii-Moriya vector lies along the plane (1st Moriya rule). Considering the three-fold rotational symmetry around $z$, we deduce that the Dzyaloshinskii-Moriya vector is necessarily along $z$. Notice that the Dzyaloshinskii-Moriya vector also possesses an in-plane component that has three-fold symmetry. We now move on to the Fe$_1$-Fe$_2$ pair, depicted on [Fig. \ref{Fig1}(c)]. Here, the same symmetry principles apply and we find that the Dzyaloshinskii-Moriya vector must be perpendicular to the Fe$_1$-Fe$_2$ segment. Notice that in the case of the Fe$_3$-Fe$_2$ pair, the Dzyaloshinskii-Moriya vector adopts the opposite orientation. Finally, the interaction between Fe$_2$-Fe$_2$ [Fig. \ref{Fig1}(d)] is similar to the one obtained for Fe$_1$-Fe$_1$ so that the Dzyaloshinskii-Moriya vector possesses a constant $z$ component and a staggered planar component.\par

To understand what is the overall influence of these different Dzyaloshinskii-Moriya vectors, one needs to remark that the Dzyaloshinskii-Moriya interactions involving either Fe$_1$ or Fe$_3$ are systematically opposite to each other because of the mirror symmetry normal to the (001) plane. Therefore, one might expect small magnetization canting at the level of the unit cell but no overall effect on the long range. This remark coincides with the absence of long-wavelength interaction emphasized in the previous section. Furthermore, the possible small canting of the magnetic moments in the unit cell could be an explanation for the topological Hall effect reported in Ref. \onlinecite{You2019}.\par

What is particularly interesting is that whereas the in-plane component of the Dzyaloshinskii-Moriya vector is staggered, the perpendicular ($z$) component of the Fe$_2$ layer remains constant over the unit cell. Therefore, one expects that at intermediate range (i.e., beyond the size of a unit cell), the atomistic Dzyaloshinskii-Moriya energy reads $E_{\rm DM}=D{\bf z}\cdot({\bf S}_1\times{\bf S}_2)$. This interaction is carried by the central Fe elements and its magnitude is therefore associated with the electrostatic environment of Fe$_2$. The latter remark is important because the only heavy element of the structure is Te, which is located further apart from Fe$_2$. Therefore, one would expect the overall magnitude of the Dzyaloshinskii-Moriya interaction to remain small.

\paragraph*{Spin spiral calculations - } To confirm the analysis provided above, we performed DFT calculations on Fe$_3$GeTe$_2$. We used the full-potential linearized augmented-plane-wave (FLAPW) method as implemented in the FLEUR software \cite{Fleur}. Applying the generalized Bloch theorem \cite{Kurz2004}, we first self-consistently compute the total energy of the system for spin spirals with different wavelengths $q$ including the scalar-relativistic effects but in the absence of spin-orbit coupling, $\epsilon^{\rm SR}(q)$. Then, we turn on the spin-orbit coupling and compute the spin spiral dispersion at the first order only, $\epsilon^{\rm SOC}(q)$. The scalar-relativistic dispersion $\epsilon^{\rm SR}(q)$ provides the magnetic exchange parameter $A$, while the difference $\epsilon^{\rm SOC}(q)-\epsilon^{\rm SR}(q)$ provides a measure of the magnetic anisotropy $K$ (at $q$=0) and Dzyaloshinskii-Moriya interaction $D$ \cite{Heide2009,Ferriani2008,Belabbes2016}.\par

For the structural relaxation, we employed the generalized gradient approximation (GGA) \cite{Perdew1996}, obtaining a relaxed lattice constant of 4.01~\AA~  for Fe$_3$GeTe$_2$ monolayer. For the magnetic calculations, we used the local density approximation (LDA) \cite{Perdew1981}. In all calculations, we selected the radii of muffi-tin spheres around 2.1 a.u for Ge and Fe, and 2.6 a.u for Te, where a.u is the Bohr radius. The linearized augmented plane-wave basis functions included all wave vectors up to $k_{max}$ = 3.8 a.u$^{-1}$ in the interstitial region and in the muffin-tin spheres, and basis functions including spherical harmonics up to $l_{max}$ = 8 were taken into account. For collinear (non-collinear + spin-orbit coupling), the calculations were performed on a dense mesh of 512 (1024) $k$-points in the full two-dimensional Brillouin zone.\par 

\begin{center}
\begin{table}
\begin{tabular}{c|c|c}
Element&Spin moment&Orbital moment\\\hline\hline
Te$_{1,2}$&-0.024&-0.02\\
Fe$_{1,3}$&2.267&0.083\\
Fe$_2$&1.287&0.19\\
Ge&-0.06&0.0065
\end{tabular}\caption{Spin and orbital moments of the various elements (in units of $\mu_B$).\label{table:1}}
\end{table}\end{center}

\begin{figure}
\begin{center}
        \includegraphics[width=6cm]{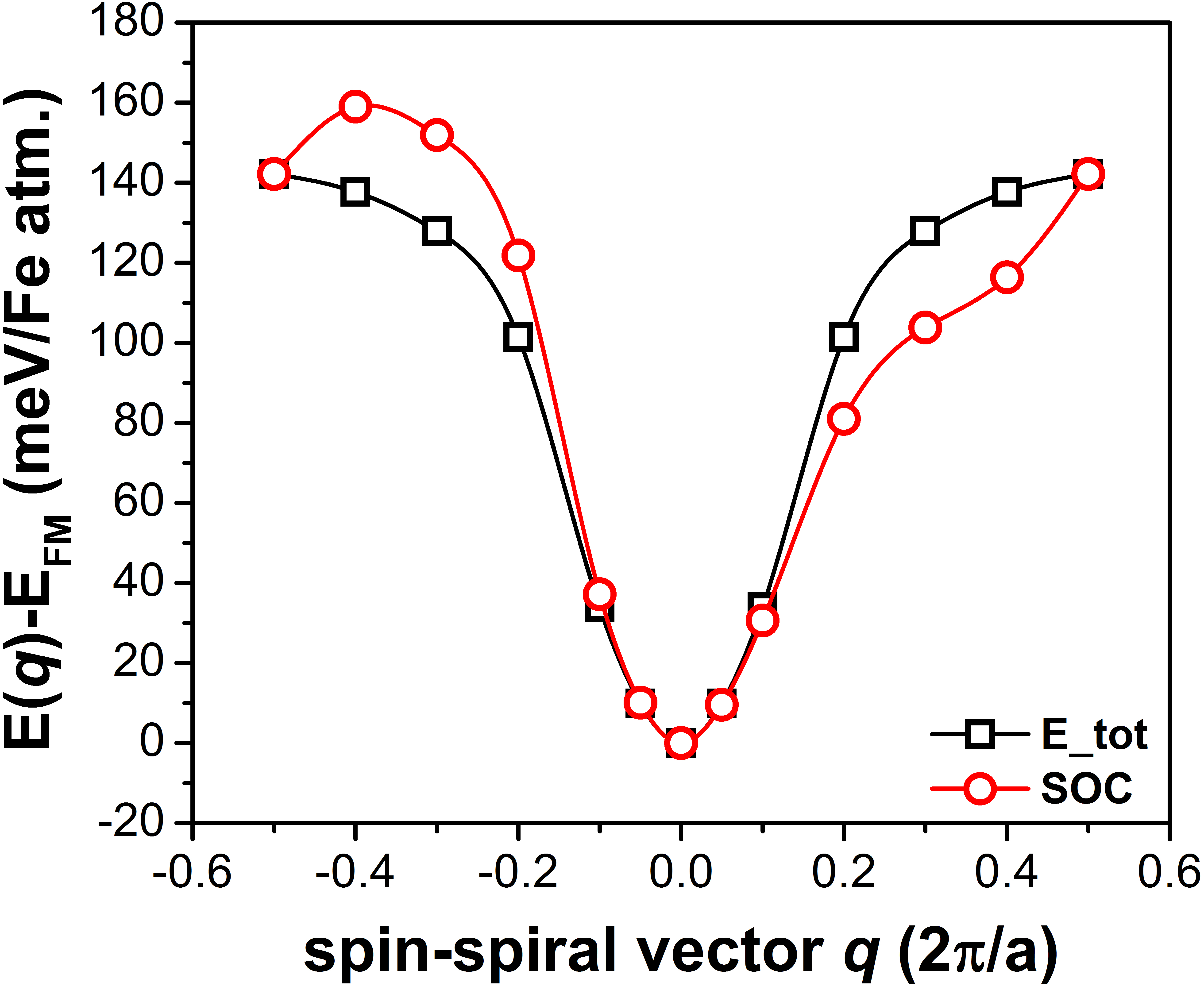}
      \caption{(Color online) Spin spiral dispersion for N\'eel in-plane along the $\Gamma-{\rm M}$ path. The vertical shift due to magnetic anisotropy has been removed manually for clarity.\label{Fig2}}
\end{center}
\end{figure}
Based on the procedure described above, we obtain the spin- and orbital-resolved magnetic moments displayed in Table \ref{table:1} and a perpendicular magnetic anisotropy $K$=1.3 meV/Fe. We have performed spin spiral dispersion calculation for the three standard spin spiral configurations: (a) N\'eel out-of-plane, (b) Bloch out-of-plane and (c) N\'eel in-plane (see Ref. \onlinecite{Belabbes2016b} for an explicit representation of these configurations). We found that only the N\'eel in-plane spin spiral displays Dzyaloshinskii-Moriya interaction, as expected from the previous analysis. For this spiral, the magnetic exchange parameter is $A$=47 meV/Fe. The corresponding spin spiral dispersion along the path $\Gamma-{\rm M}$ is represented on Fig. \ref{Fig2}, with (red) and without (black) spin-orbit coupling. One clearly sees that turning on spin-orbit coupling distorts the spin spiral dispersion in an antisymmetric manner. In contrast, neither N\'eel nor Bloch out-of-plane spin spirals display any antisymmetric contribution upon turning on the spin-orbit coupling (not shown), indicating that the Dzyaloshinskii-Moriya vector does not possess in-plane components.\par

\begin{figure}
\begin{center}
        \includegraphics[width=9cm]{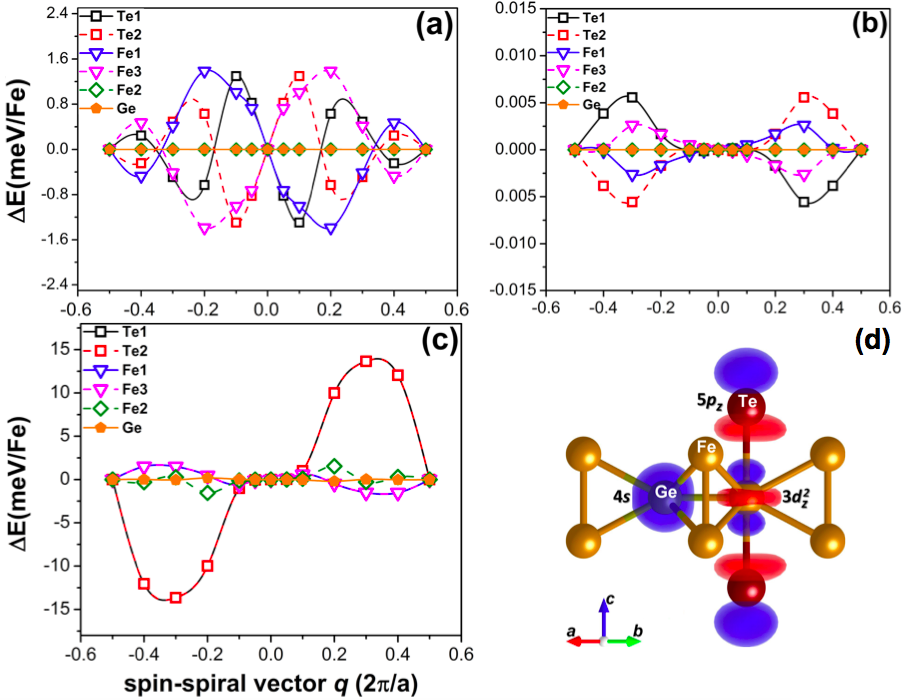}
      \caption{(Color online) Contribution of the different elements to the antisymmetric part of the spin spiral dispersion for (a) N\'eel out-of-plane, (b) Bloch out-of-plane and (c) N\'eel in-plane configurations. (d) Kohn-Sham orbitals at $\Gamma$-point showing the strong Te 5$p_z$-Fe$_2$ 3$d_{z^2}$ hybridization responsible for the large perpendicular orbital moment on Fe$_2$.  \label{Fig3}}
\end{center}
\end{figure}

In order to given a complete picture of the microscopic origin of the perpendicular Dzyaloshinskii-Moriya interaction, we analyzed the contribution of the various elements (Fe$_{1,3}$, Fe$_2$, Te and Ge) on the antisymmetric spin spiral dispersion for the three configurations. The results are displayed on Fig. \ref{Fig3}. In the case of N\'eel out-of-plane spin spiral [Fig. \ref{Fig3}(a)], the main contributions come from top and bottom Fe$_{1,3}$ as well as from the top and bottom Te elements. Interestingly, these contributions are sizable in magnitude but cancel each other by symmetry. In the case of Bloch out-of-plane spin spiral [Fig. \ref{Fig3}(b)], we obtain the same cancellation by symmetry but the magnitude of the individual contributions remain extremely small (a few $\mu$eV/Fe). Such a small magnitude is below the accuracy of our calculations and is therefore insignificant. Finally, for the N\'eel in-plane spin spiral [Fig. \ref{Fig3}(b)], the antisymmetric dispersion is dominated by the hybridization between the 5$p_z$ orbitals of Te and the 3$d_{z^2}$ orbitals of the central Fe$_2$ element [see Fig. \ref{Fig3}(b)]. This hybridization induces a large orbital momentum on Fe$_2$ (0.19$\mu_B$), as displayed in Table \ref{table:1}, that is responsible for the observed Dzyaloshinskii-Moriya interaction. In contrast, the Fe$_{1,3}$ elements have a much smaller contribution. In fact, although they also experience a perpendicular Dzyaloshinskii-Moriya vector [see Fig. \ref{Fig1}(b)], they chemical environment does not promote a large orbital moment (0.083$\mu_B$) and therefore yields a small Dzyaloshinskii-Moriya interaction.\par

\begin{figure}[h!]
\begin{center}
        \includegraphics[width=6cm]{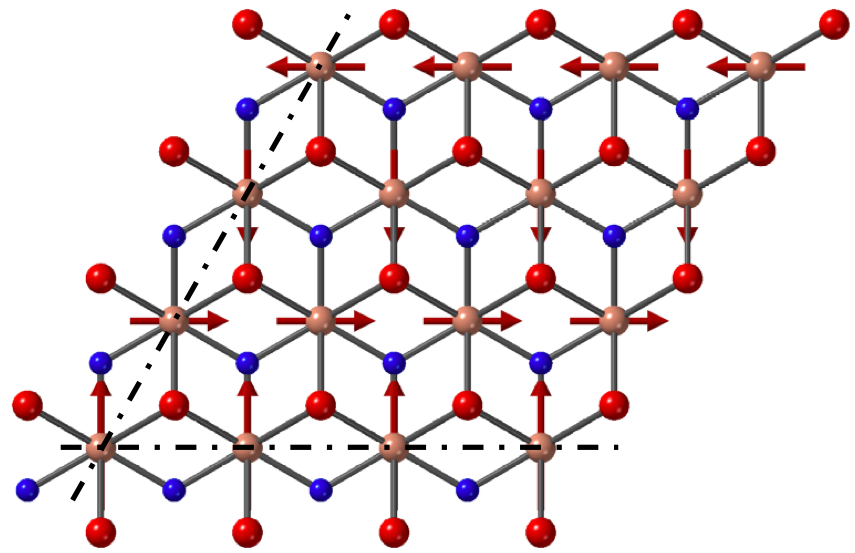}
      \caption{(Color online) Example of a planar homochiral spin spiral propagating along the low symmetry (100) direction. This six-fold degenerate direction, identified by the dashed line, is characterized by mirror symmetry breaking, enabling perpendicular Dzyaloshinskii-Moriya interaction. \label{Fig4}}
\end{center}
\end{figure}

We complete this analysis by discussing the potential influence of the perpendicular Dzyaloshinskii-Moriya interaction of the stabilization of magnetic textures. It is clear that the large perpendicular magnetic anisotropy of Fe$_3$GeTe$_2$ ($K$=1.3 meV/Fe) hinders the stabilization of N\'eel in-plane spin spirals. Nevertheless, for the sake of the discussion let us disregard the role of the magnetic anisotropy and only focus on the influence of the Dzyaloshinskii-Moriya interaction itself. What is particularly remarkable is that the antisymmetric dispersion is quite different from the dispersion obtained at, e.g., transition metal interfaces \cite{Ferriani2008,Kashid2014,Belabbes2016}. In the latter, the antisymmetric contribution of the dispersion has a large slope around $q=0$, from which the long-wavelength Dzyaloshinskii-Moriya coefficient is usually extracted. In Fig. \ref{Fig2}, the slope close to $q=0$ vanishes and the antisymmetric dispersion only takes off aways from the origin. This feature means that the Dzyaloshinskii-Moriya interaction has no impact in the long wavelength limit and is unlikely to stabilize large ($>$10 nm) chiral textures. Nonetheless, it does tend to stabilize short-wavelength spin spirals. Indeed, the dispersion is peaked around $q\approx\frac{2\pi}{\sqrt{3}a}$. Considering that this dispersion is computed along the $\Gamma-{\rm M}$ path, it means that Dzyaloshinskii-Moriya interaction tends to stabilize planar homochiral spin spirals propagating along a low symmetry direction of the Fe$_3$GeTe$_2$ crystal. Figure \ref{Fig4} shows such a planar spin spiral extended along the (100) direction of the crystal (dashed lines). The (100) direction is indeed characterized by broken mirror symmetry and therefore favors perpendicular Dzyaloshinskii-Moriya interaction, consistently with the analysis provided above.\par

\paragraph*{Conclusion - }
Using both symmetry arguments and first principle calculations, we have shown that the Dzyaloshinskii-Moriya interaction of Fe$_3$GeTe$_2$ adopts the form $D{\bf z}\cdot({\bf S}_1\times{\bf S}_2)$, with a Dzyaloshinskii-Moriya vector perpendicular to the (001) plane. This interaction is unable to stabilize the N\'eel skyrmions reported recently in thick Fe$_3$GeTe$_2$ layers, but it possesses remarkable characteristics. It vanishes in the long wavelength limit and only survives for small textures as it tends to stabilize planar spin spiral with wavevector $q\approx\frac{2\pi}{\sqrt{3}a}$ and propagating along (100) direction, i.e., perpendicular to the (110) mirror plane direction. Nonetheless, in realistic situations, the large perpendicular magnetic anisotropy of Fe$_3$GeTe$_2$ monolayers prevents the formation of such planar spin spirals, at least in the monolayer limit. Cancelling this perpendicular anisotropy by surface engineering represents an appealing challenge both experimentally and theoretically as it could open avenues for the generation of unusual chiral textures.

\begin{acknowledgments}
S.L. and A.M. thank G. Bihlmayer and S. Bl\"ugel for useful discussions. This work was supported by the King Abdullah University of Science and Technology (KAUST) through the Office of Sponsored Research (OSR) [Grant Number OSR-2017-CRG6-3390]. K.-W. K acknowledges support from the KIST institutional program and the National Research Foundation of Korea (2020R1C1C1012664) 
\end{acknowledgments}


\begin{thebibliography}{40}
\bibitem{Huang2017b} Huang et al., Nature {\bf546}, 270 (2017).
\bibitem{Bonilla2018} Bonilla et al., Nature Nanotechnology {\bf13}, 289 (2018).
\bibitem{Li2018f} Li et al., Advanced Materials {\bf30}, 1801043 (2018).
\bibitem{Sun2019} Sun et al., arXiV:1909.09797v1 (2019).
\bibitem{Deng2018} Deng et al., Nature {\bf563}, 94 (2018).
\bibitem{Li2019d} Li et al., ACS Applied Materials \& Interfaces {\bf11}, 10729 (2019).
\bibitem{Zhang2020a} Zhang et al., Applied Physics Letters {\bf116}, 042402 (2020).
\bibitem{Park2020} Park et al., Nano Letters {\bf20}, 95 (2020).
\bibitem{Alghamdi2019} Alghamdi et al., Nano Letters {\bf19}, 4400 (2019).
\bibitem{Wang2019c} X. Wang, J. Tang, X. Xia, C. He, J. Zhang, and Y. Liu, Science Advances {\bf5}, eaaw8904 (2019).
\bibitem{Xu2019} J. Xu, W. A. Phelan, and C.-l. Chien, Nano Letters {\bf19}, 8250 (2019).
\bibitem{Fang2019} Fang et al., Applied Physics Letters {\bf115}, 212402 (2019).
\bibitem{Wang2018c} Z. Wang, D. Sapkota, T. Taniguchi, K. Watanabe, D. Mandrus, and A. F. Morpurgo, Nano Letters {\bf18}, 4303 (2018).
\bibitem{Albarakati2019} Albarakati et al., Science Advances {\bf5}, eaaw0409 (2019).
\bibitem{Wang2019d} Wang et al., arXiv:1907.08382 (2019).
\bibitem{Ding2019} Ding et al., arXiv:1912.11228 (2019).
\bibitem{Park2019} Park et al., arXiv:1907.01425 (2019).
\bibitem{Dzyaloshinskii1957} I. Dzyaloshinskii, Soviet Physics JETP {\bf5}, 1259 (1957).
\bibitem{Moriya1960} T. Moriya, Physical Review {\bf120}, 91 (1960).
\bibitem{Nagaosa2013} N. Nagaosa and Y. Tokura, Nature nanotechnology {\bf8}, 899 (2013).
\bibitem{Chen2015b} G. Chen, A. Mascaraque, A. T. N?Diaye, and A. K. Schmid, Applied Physics Letters {\bf106}, 242404 (2015).
\bibitem{Johansen2019} O. Johansen, V. Risinggard, A. Sudbo, J. Linder, and A. Brataas, Physical Review Letters {\bf122}, 217203 (2019).
\bibitem{Freimuth2014} F. Freimuth,S. Bl\"ugel, and Y. Mokrousov, Journal of Physics: Condensed Matter {\bf26}, 104202 (2014).
\bibitem{Zelezny2017} Zelezny et al., Physical Review B {\bf95}, 014403 (2017).
\bibitem{Raeliarijaona2018} A. Raeliarijaona, R. Nepal and A. A. Kovalev, Physical Review Materials {\bf2}, 124401 (2018).
\bibitem{Hals2017} K. M. D. Hals and K. Everschor-Sitte, Physical Review Letters {\bf119}, 127203 (2017).
\bibitem{Rohart2013} S. Rohart and A. Thiaville, Physical Review B {\bf88}, 184422 (2013).
\bibitem{You2019} You et al., Physical Review B {\bf100}, 134441 (2019).
\bibitem{Fleur} http://www.flapw.de
\bibitem{Kurz2004} P. Kurz, F. F\"orster, L. Nordstr\"om, G. Bihlmayer, and S. Bl\"ugel, Physical Review B {\bf69}, 024415 (2004).
\bibitem{Heide2009} M. Heide, G. Bihlmayer, and S. Bl\"ugel, Physica B: Condensed Matter {\bf404}, 2678 (2009).
\bibitem{Ferriani2008} Ferriani et al., Physical Review Letters {\bf101}, 027201 (2008).
\bibitem{Belabbes2016} A. Belabbes, G. Bihlmayer, F. Bechstedt, S. Bl\"ugel, and A. Manchon, Physical Review Letters {\bf117}, 247202 (2016).
\bibitem{Perdew1996} J. P. Perdew, K. Burke, and M. Ernzerhof, Physical Review Letters {\bf77}, 3865 (1996).
\bibitem{Perdew1981} J. P. Perdew and A. Zunger, Physical Review B {\bf23}, 5048 (1981).
\bibitem{Belabbes2016} A. Belabbes, G. Bihlmayer, S. Bl\"ugel, and A. Manchon, Scientific Reports {\bf6}, 24634 (2016).
\bibitem{Kashid2014} Kashid et al., Physical Review B {\bf90}, 054412 (2014).
\end{thebibliography}
\end{document}